# On challenges and opportunities of designing integrated IT platforms for supporting knowledge works in organizations

Arijit Laha
Center for Knowledge Driven Information Systems
Software Engineering and Technology Labs
INFOSYS TECHNOLOGIES LTD
Hyderabad
E-mail: Arijit_laha@infosys.com

**ABSTRACT**

Designing and implementing comprehensive IT-based support environments for KM in organizations is fraught with many problems. Solving them requires intimate knowledge about the *information usage in knowledge works* and the scopes of technology intervention. In this paper, the Task-oriented Organizational Knowledge Management or TOKM, a design theory for building integrated IT platforms for supporting organizational KM, is proposed. TOKM brings together two apparently mutually exclusive practices of building KM systems, the task-based approach and the generic or universalistic approach. In developing the design, the information requirements of knowledge workers in light of an information usage model of knowledge works is studied. Then the model is extended to study possibilities of more advanced IT support and formulate them in form of a set of meta-requirements. Following the IS design theory paradigm, a set of artifacts are hypothesized to meet the requirements. Finally, a design method, as a possible approach of building an IT-based integrated platform, the Knowledge Work Support Platform (KWSP) to realize the artifacts in order to meet the requirements, is outlined. The KWSP is a powerful platform for building and maintaining a number of task-type specific Knowledge Work Support Systems (KWSS) on a common sharable platform. Each KWSS, for the task-type supported by it, can be easily designed to provide extensive and sophisticated support to individual as well as group of knowledge workers in performing their respective knowledge work instances.

## INTRODUCTION

Computer science, or more specifically the computer-based information technology (IT), facilitates working with *information* at a scale and efficiency level unprecedented in human history. Naturally, IT is recognized as the core technology enabler of knowledge management (KM) efforts in modern organization (Alavi and Leidner 1999). IT-based support environments for KM, popularly known as KMS, have complex multi-faceted nature involving not only technological aspects, but also social, cultural and behavioral aspects of the organizations and their workers (Alavi and Leidner 1999). This is a crucial and very useful insight into the problem. It draws attention to the fact that the solution is beyond *technology alone*. In general, when people are introduced to powerful technology, a complex and emergent interplay of peoples' behavioral aspects and functionalities provided by the technologies take place. This has been subject of studies by many IS researchers (Markus 1994, Ruhleder 1994, Lapointe and Rivard 2005, Burton-Jones and 2007). Similar studies on usage of KMS, where human actors need to exercise their intellectual and cognitive abilities, have also been reported by several researchers (Schultze 2000, Becerra-Fernandez and Sabherwal 2001, Stenmark and Lindgren 2006).

Today a large number of IT tools and technologies are viewed as part of KMS. But most of them owe their origin to requirements different from those of KM. The ubiquitous e-mail is an electronic version of age-old physical mailing system. The "document management systems", though evolved significantly (Dourish et al. 2000) over time, their core functionalities are aimed at fulfilling organizational legal and/or regulatory requirements for preserving documents. The "content management systems" (McKeever 2003) evolved out of document management systems in order to accommodate the demands for systematically managing contents of large websites. Various web-based technologies, though potentially very useful for KMS, are usually designed to cater a much larger gamut of generic requirements. The collaboration (CSCW) tools (Rama and Bishop 2006) evolved out of the requirements of joint document preparation; typically do not systematically facilitate knowledge work requirements such as argumentation and validation of arguments. This situation leads many IT researchers and practitioners to adopt a "mix-n-match" approach of repackaging various functionalities of available tools in designing the IT support environments for KM (Marwick 2001, Hung et al. 2007), which are often found inadequate. The problem is pointed out by Maier (2005, p. 429) as "… the solution is still not there and many businesses trying to implement these technologies have been frustrated by the fact that the technologies certainly could not live up to the overly high expectations".

To overcome the problem, we need to build a deeper understanding of the KM issues that can help *designing* and *implementing* KMS with a greater possibility of success. It must be remembered that IT deals with information, i.e., electronically encoded "messages". IT-systems offer various *predefined* functionalities to work with them in "algorithmic" and/or "probabilistic" manner (Saracevic 1999). Thus, using IT for supporting KM, where human actors' intellectual abilities play most crucial roles, is far from straightforward. From the IT perspective, the KM challenge is to enable the human actors/experts, also known as the knowledge workers, to perform their tasks with high degree of efficiency and accuracy. The possible role of IT in KM can be stated with deceptive simplicity: enabling human actors/experts working *efficiently* with *relevant* information. Here the key issues are the "relevance" of the information received by the knowledge workers and "efficiency" of the system in delivering them. Thus, in order to build a deeper insight regarding how to meet these requirements as best as possible by using IT, we need to understand clearly the nature of *information usage* in knowledge works.

# CONTRIBUTION OF THE PAPER

In pursuance of the objective stated above, in this paper I propose a model of *information usage in knowledge works* which allows us to study various processes, systemic as well as cognitive, involved in information usage and conventional ways of using IT to support them. Then I extend this model to explore broader scope of IT interventions for improving the information usage capabilities of knowledge workers. Finally I propose a generic IT system architecture, the Knowledge Work Support Platform (KWSP) which can support building, using and maintenance of a set of task-specific but interoperating Knowledge Work Support Systems (KWSS) (Burstein and Linger 2003, Stenmark and Lindgren 2006). The KWSSes, leveraging the support provided by the KWSP can fulfill the requirement of improved IT intervention. The content of the paper is organized as an IS design theory (Walls et al. 1992), the Task-oriented Organizational Knowledge Management (TOKM).

# PRELIMINARIES

The theory, TOKM, deals at its core with the issues related to *information usage in knowledge works*, i.e., how does human knowledge workers use information and how the related processes can be supported effectively by technology interventions. The concepts I shall build upon, though might have originated in myriads of disciplines, I shall mostly draw upon them as available through the literature of the three fields of studies related in different ways to *information*, namely the *information systems*, the *information science* and the *information technology*. Readers interested in deeper exploration of the concepts/issues can use the "references" sections of the referred articles to further their quest.

## *Handling Information*

All the three fields, information system, information science and information technology have "information" as their central themes. Nevertheless, despite significant overlaps, each of these fields has its own distinct perspective, goal and even lexicon, which render subtle but significantly different semantic to same or similar terms. In the following we shall discuss various aspects of each of them relevant to our discourse.

### The "Information Systems"

In IS literature there are a number of views on the fundamental concepts "knowledge" and "information" (Nonaka 1994, Tuomi 1999, Alavi and Leidner 2001). The ideas of "tacit knowledge" and "explicit knowledge" introduced by Polanyi (1967) and later interpreted by Nonaka (1994) in context of knowledge creation in organizations have influenced several authors (Zack 1999, Duffy 1999, Tiwana 2000).

For the purpose of the current paper I shall adopt the view of Alavi and Leidner (2001) regarding knowledge and information that "knowledge" resides in the mind of individual human beings, the "knowers"; its possession is manifested by the individuals' capabilities of performing works. In society, it often becomes imperative for people to work together as well as to acquire capabilities of performing new works. This necessitates sharing and/or communication of knowledge among people. Unfortunately, there exists *absolutely* no direct means of achieving that. It is mediated by creation and absorption of information by human actors, i.e., "… information is converted to knowledge once it is processed in the mind of individuals and knowledge becomes information once it is articulated and presented in the form of text, graphics, words, or other symbolic forms, … knowledge does not exist outside of an agent (a knower): it is indelibly shaped by one's needs as well as one's initial stock of knowledge, …individuals to arrive at the same understanding of data or information, they must share a certain knowledge base" (Alavi and Leidner 2001, p. 109). The scenario is depicted graphically in Figure 1.

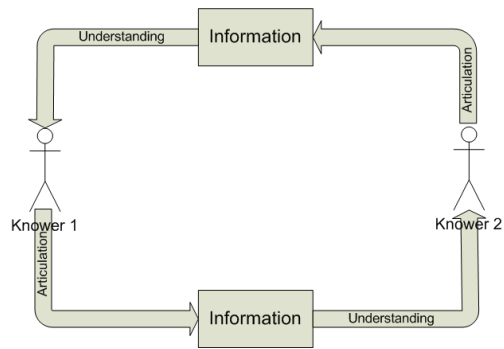

Figure 1: Knowledge sharing/communication

## The "Information Science"

Information science is defined as "… the study of the gathering, organizing, storing, retrieving, and dissemination of information" (Bates 1999, p. 1044). Over time, the field has grown into a large and multi-disciplinary one. Bates (1999) provides an excellent account of linkage of information science with various disciplines. From a more functional viewpoint, "… information science is a field of professional practice *and* scientific inquiry addressing the problem of effective communication of knowledge records— "literature"—among humans in the context of social, organizational, and individual need for and use of information. The key orientation here is the problem *of need for and use of information, as involving knowledge records*. To provide for that need, information science deals with specifically oriented information techniques, procedures, and systems" (Saracevic 1999, p. 1055). Saracevic identifies three central ideas of information science (Saracevic 1999, p. 1052) as (1) information retrieval (IR): "…providing for processing of information based on formal logic", (2) relevance: "… directly orienting and associating the process [of IR] with human information needs and assessments" and (3) Interaction: "… enabling direct exchanges and feedback between systems and people engaged in IR processes" .These ideas form the areas of use of IT in context of information science.

Of the three ideas, the relevance is most complex due to its inherent subjectivity (Meadows 2008). A detailed account of the literature on relevance can be found in (Mizzaro 1997). The issue of "judgment of relevance" has been studied by many researchers (Spink et al. 1998, Tang and Solomon 1998, Anderson 2006). In an effort to provide a relatively unified view of relevance, Mizzaro (1997 p. 811) defines relevance as *a relation between a pair of entities*, drawn one each from two sets of entities, where the first set is consisted of (1) *document*: the physical entity that the user of an Information Retrieval System will obtain after his seeking of information; (2) *surrogate*: a representation of a document. It may assume different forms and may be made up by one or more of the following: Title, list of keywords, author(s) name(s), bibliographic data (date and place of publication, publisher, pages, and so on) , abstract, extract (sentences from the document) , and so on; and (3) *information*: what the user receives when reading a document.

The second set contains (1) *problem*: that which a human being is facing and that requires information for being solved; (2) *information need*: a representation of the problem in the mind of the user. It differs from the problem because the user might not perceive in the correct way his problem; (3) *request*: a representation of the information need of the user in a ''human'' language, usually in natural language; and (4) *query*: a representation of the information need in a ''system'' language, for instance Boolean. Also, each of the entities can be decomposed into (1) *topic*: that which refers to the subject area to which the user is interested; (2) *task*: that which refers to the activity that the user will execute with the retrieved documents; and (3) *context*: that which includes everything not pertaining to topic and task, but however affecting the way the search takes place and the evaluation of results.

## The "Information Technology"

In modern era, computer-based tools and technologies, known collectively as the "Information Technologies" (IT), or sometimes as the "Information and Communication Technologies" (ICT), play a crucial role in supporting people dealing with information. However, despite the power of IT vis-à-vis processing, manipulating and storing information, its perspective of information is fairly narrow. IT, in itself, treats information as "signals" or "messages" and, work with it *algorithmically* or *probabilistically* without any cognitive elements (Saracevic 1999). Thus, how to use its immense power effectively in context of problems beyond those can be specified adequately in terms of algorithmic and/or probabilistic processing, require careful examination of the requirements of those problems in terms of *information usage*.

## *The "Knowledge Management"*

The various aspects of the problem of knowledge management has been studied many researchers (Wiig 1993, Drucker 1993, Nonaka 1994, O'Dell 1996, Prusak 1997, Davenport and Prusak 1998, Davenport 2005). From a user-centric perspective Markus (2001) developed a theory of knowledge reuse in terms four categories of users, where KM is interpreted as means for enabling these classes of users. From IS perspective, Alavi and Leidner (2001) group the processes involved in knowledge management into four categories, *creation, storage/retrieval, transfer* and *application*. Following the categorization, they studied the mechanics of supporting them using technology in socially situated environment.

Approaches to implementing KM in organizations are broadly divided into categories, the process/task approach and the infrastructure/generic approach (Jennex and Olfman 2005). While an all-embracing generic approach is closer in concept to what an organization may ultimately need to accomplish, the requirements of addressing non-technological issues may put substantial hurdles in the path of realizing it. The root problem is the social, cultural and behavioral diversity among various units and subunits of the organization (Pisano 1994). This often results in serious reduction of effectiveness of such *universalistic* (Becerra-Fernandez and Sabherwal 2001) or top-down approaches.

## Task-based Knowledge Management (TbKM)

On the other hand, the task-based approaches focus on individual tasks and the workers performing them. In recent years, a host of extensive studies (Schultze 2000, Becerra-Fernandez and Sabherwal 2001, Stenmark and Lindgren 2006, Burstein and Linger 2003, Aarons et al. 2006) along with other crucial findings, point out the necessity of building information systems supporting the KM, popularly known as the "Knowledge Management Systems" or KMS adapted to catering the needs of *knowledge workers* in ways which directly affect their performance of *knowledge-intensive tasks*. Such approaches of knowledge management are known as the *task-based knowledge management* or TbKM (Leake et al. 1999, Burstein and Linger 2003).

Burstein and Linger (2003), based on their extensive field works, proposed "Task-based Knowledge Management" or TbKM as a framework for studying and analyzing the characteristics of knowledge-intensive tasks. They define a "task" as a substantially invariant activity with outcomes, including tangible outputs. A task is performed by socially situated "actors". They use the term "knowledge work" referring to the collection of activities that constitute a task. This clearly makes a huge shift of paradigm by changing the focus to *task* instead of *organization*. As a consequence, a designer of KMS, instead of studying the KM requirement of whole organization in its all glorious diversity, can study the requirements of individual workers or community of workers (CoP) (Wenger 1998) involved in performing specific classes of tasks and design system supports adapted to the task-specific characteristics.

An information system designed for supporting a targeted task is called a *Knowledge work Support System* or KWSS (Burstein and Linger 2003, Stenmark and Lindgren 2006), capable of facilitating *thinking* and *doing* in a integrated manner (Aarons et al. 2006). The TbKM is a framework for developing KWSSes, as stand-alone or independent systems, each supporting *performance of instances* of a single targeted task or task type. Burstein and her coworkers demonstrated the power of TbKM in context of systems studied for several complex tasks (Burstein and Linger 2003), e.g., the weather forecasting task in Australian department of meteorology (Aarons et al. 2005).

## The *Integrated* Task-based Knowledge Management

From a holistic perspective, there are a number of knowledge-intensive tasks in organizations whose proper executions have crucial effect on the well-being, competitiveness, even mere survival of the organizations. There are often complex *many-to-many interdependencies* among the tasks. For example, while planning a marketing campaign; one worker may want to use result of other tasks, different both from type/class as well as instance/episode perspective, like organizational policy, budget, and market research reports etc. Clearly, to cater such requirement scenarios, building stand-alone KWSSes for different tasks may not be an acceptable approach of implementing KM initiative in the organization. Such conflicting requirements scenario is not uncommon in IT. Johnson and Nardi (1996) conducted a study on user preference of task-specific versus generic software applications. Their findings were very much in line of the above.

I contend that a viable solution out of this dichotomy is to build an integrated technology platform, the "Knowledge Work Support Platforms" (KWSP), on top of which multiple KWSSes supporting a range of tasks can be built. The platform will allow building one KWSS at a time, thus keeping the complexity manageable and enabling the designer to use established frameworks such as TbKM for the purpose. On the other hand, the platform will allow KWSSes to systematically inter-operate.

## *Presentation of TOKM*

In field of IS, use of design theories, known as *IS design theories* (ISDT), were introduces and formalized by Walls et al. (1992). They defined ISDT as "… a prescriptive theory based on theoretical underpinnings which say how a design process can be carried out in a way which is both effective and feasible" (Walls et al. 1992, p. 37). An ISDT is used for building *class of systems/artifacts* addressing a *class of problems* (as opposed to a tool addressing a single problem) (Walls et al. 1992). Works following the ISDT research paradigm now form a major branch of IS research (Hevner et al. 2004). Many of the influential theories in IT/IS, e.g. relational database theory (Codd 1970) and software development life cycle management (SDLC) theory, are prime examples design theories (Walls et al. 1992). In the current paper, I shall model the proposed theory, the Task-oriented Organizational Knowledge Management or TOKM as an IS design theory.

A design theory or simply, a design is both *a product* as well as *a process* (Walls et al. 1992). Walls et al. (1992, p. 42) assert that "…as a product, a design is "a plan of something to be done or produced": as a process, to design is "to so plan and proportion the parts of a machine or structure that all requirements will be satisfied"". According to this viewpoint they identified various components of an ISDT as shown in Table 1. The interrelationship among these components of ISDT and their correspondence with elements of TOKM is depicted in Figure 2.

| Class | Component | Content |
| --- | --- | --- |
| Design Product | Meta-requirements | Describes the class of goals to which the theory applies. |
| | Meta-design | Describes a class of artifacts hypothesized to meet the meta-requirements |
| | Kernel theories | Theories from natural or social sciences governing design requirements. |
| | Testable design product hypotheses | Used to test whether the meta-design satisfies the meta-requirements. |
| Design Process | Design method | A description of procedure(s) for artifact construction |
| | Kernel theories | Theories from natural or social sciences governing design process itself |
| | Testable design process hypotheses | Used to verify whether the design method results in an artifact which hypotheses is consistent with the meta-design. |

Table 1: Components of an ISDT (Walls et al. 1992)

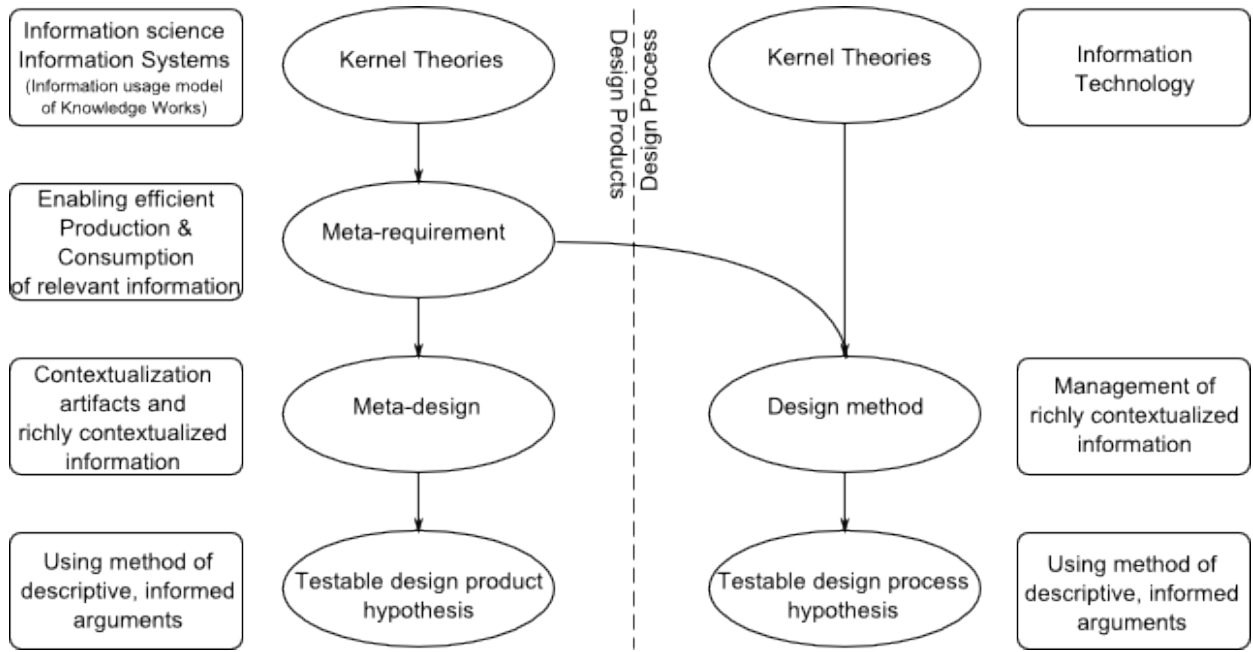

Figure 2: Correspondence between TOKM elements and ISDT components

**AN INFORMATION USAGE MODEL OF KNOWLEDGE WORK: THE KERNEL THEORY AS DESIGN PRODUCT**

In the current approach, the idea of *knowledge work* (KW) is central one. Markus et al. (2002) proposed an IS design theory for building information systems supporting *emergent knowledge processes* (EKP). Typical EKPs include basic research, new product development, strategic business planning, organization design etc. Evidently, the EKPs are prime examples of knowledge-intensive tasks or knowledge works involving human creativity and reflectivity as well as extensive and complex information requirements. The KM initiative in any organization must provide enough support for them. However, such systems can be easily recognized as process/task focused KMSes In an organizational context, as observed earlier, we have to be able to tackle a scenario where tasks have *many-to-many* relationships with respect to their *information usage pattern*, and the organization should be capable of building them easily and incrementally using re-usable components.

### *The knowledge work and knowledge worker*

It can be easily recognized that the ultimate goal of knowledge management is to *support knowledge workers in efficiently performing their work, i.e., knowledge works, in order to produce high quality results.* Quite a few different views on "knowledge work" and "knowledge workers" can be found in literature (Davis 1999, Schultze 2000, Maier 2005, Davenport 2005, Stenmark and Lindgren 2006). Each view focuses on some aspects of them relevant to the particular discourse. Further, there are often some differences in the nomenclatures used. Here, based on the relevant literature we consolidate our view on these entities and define some nomenclature to be adopted in rest of the paper.

## The "Knowledge Work"

Knowledge works are hugely complex and dynamic interplay of human mind and environment. They are essentially embodiment of *problem-solving efforts by human actors*. These are among the most complex intellectual functions (Newell and Simon 1972). The study of human problem solving has a long history and largely responsible for creation of the field of study "cognitive science". Most of the empirical studies in problem-solving have been carried out in restricted laboratory setups mostly involving puzzle-solving and theorem proving types of tasks (Langley and Rogers 2005). Such problems, however complex they may be, typically falls in the category of "well-structured problems" (Simon 1973). In contrast, most of the real-life problems lack in structure, transparency, clarity on desired solution and many other aspects. In short, they exhibit the characteristics of "ill-structured problems" (Simon 1973), which are not amenable to analysis from "information processing system" perspective (Simon 1978), which make it, if not impossible, very difficult to build AI-based automated systems for performing them. Such tasks demand significant creative and reflective capabilities (Johnson and Carruthers 2006) requiring active involvement and intellectual contribution of human actors, popularly known as the "knowledge workers".

In context of TbKM, Burstein et al (2003) define a task or knowledge work as *a system of activities* with tangible outcomes, performed by socially situated actors. This points out two aspects of tasks, namely, (1) tasks are composed of smaller units of work, the "activities" and (2) there exists an internal order within a task consisted of the activities and their interdependency, which they call the structure and process of the task. This *structural* view of knowledge works has profound implications for our theory. It follows that any given time a worker is involved in performing an *activity* as part of a *task*, a larger unit of work. Thus, in TOKM we adopt the view that *effective informational support to the knowledge workers must cater to their requirements at the activity level*.

## The "knowledge worker"

Knowledge is a complex entity of personal and tacit nature, possessed by individuals. Individuals acquire it through education, training, work-experience, interaction with other individuals as well as non-human animate and inanimate objects comprising the environment (Polanyi 1967, Prusak 1997). Possession of knowledge by an individual is manifested by his/her ability of "acting". The qualification of one's knowledge as *relevant* is meaningful only in context of specific tasks. It means that the person, within his/her pool of knowledge also possesses a part which he/she can utilize to perform those tasks.

Let us define a "knowledge worker" as one who is in possession of *relevant knowledge* in form of expertise, experience and other problem-solving abilities and is accustomed to applying them to *understand problem environments* and *formulating suitable solutions*. This practice of problem solving by human actors is known as performance of "knowledge works" or "knowledge-intensive tasks" or as we shall often use the term for brevity, simply "task". If we examine the above definition closely, we can identify three distinct concepts, (1) relevant knowledge, (2) understanding problem environment and (3) formulating solution to the problem. Let us examine them little more closely.

Even for a worker experienced in performing a certain type of task, every *instance* of the task presents significant novelty. The novelty arises due to change of various general as well as situational aspects of the environment within which the task instances need to be performed. We call such variable aspects together the "problem environment". Without proper understanding of the environment, the worker can not apply his/her knowledge. Clearly, the worker needs to acquire *useful information* from various sources, understand and synthesize them in order to build a coherent, actionable understanding of the environment. Development of this understanding takes place in the worker's mind through creation of new knowledge and/or update of the existing knowledge (Nonaka 1994). Further, the problem environment is not static. It shows emergent/dynamic behavior based on the progress of the worker as well as changes in other external factors.

A task has some tangible outcomes/deliverables (Burstein and Linger 2003), i.e., once the worker is in possession of enough knowledge so that he/she is able to formulate the solution, to make the solution usable, it needs to be worked upon/implemented. In other words, the knowledge of the solution needs to be *articulated* by means of *actions* and/or *artifacts*. The articulation of knowledge is production of *information*, which can be captured, archived, shared and manipulated (Alavi and Leidner 2001).

The above is summarized in the definition of "knowledge work" as *production and reproduction of information and knowledge* (Stehr 1994). "Production/reproduction of knowledge", as noted earlier, happens in the mind of an individual when he/she *access, process* and *understand* information. On the other hand, "Production/reproduction of information" takes place when the worker *articulates* part of his/her knowledge. From IS perspective, we can interpret the above as a **dual role** played by the workers while engaged in knowledge works. They play the role of **information consumer** while using information to create knowledge. At the same time they are **information producer** as they generate information by means of articulation of knowledge.

Following the above discussion, in this paper, I adopt the nomenclature that "tasks" are the larger unit of knowledge work, while "activities" form the part of tasks and smaller units of knowledge work. The term "knowledge work" will be used generically to mean any or both, depending on the context, of the two entities task and activity. Further, we shall use the terms "knowledge worker", "worker", "actor" and "user" interchangeably.

## *An "information usage" model of knowledge work*

From the consumption perspective, a knowledge worker faces the problem of *seeking and finding relevant information*. Similarly, as a producer the worker needs to *articulate* his/her knowledge to *produce* information in a form commensurate with the requirement of the consumers. The crux of the problem lies in the concept of "relevance". Let us examine the problem in context of knowledge works.

### "Relevance" in knowledge works

Let us examine the roles of the ten entities (Mizzaro 1997) identified earlier in information seeking with respect to knowledge works. While the documents serve as containers of information and surrogates help the IR system (to speed-up search using indexes built on surrogates) as well as the workers (allowing them to quickly assess the content of documents without actually reading them), the workers ultimately need *information*. To find the required information, a knowledge worker first needs to *perceive* the nature of the information he/she needs. This *perceived information needs*, in turn, depends on his/her perception of entities such as the *problem*, the *domain* (i.e., topic in Mizzaro (1997)) and the *task* as well as *activity* at-hand as part of the larger task. The information need, once perceived, to be satisfied with help of IR system, need to be *articulated* in form of natural language *request* (if the IR system has interface to handle that) and/or *queries* (in a language amenable to processing by the IR system). Based on the *articulated information needs*, the IR system *searches* the archive, *retrieves* and presents to the user a set of information (actually the containing documents), which the user subject to "judgment of relevance", based on the perceived information needs. The portion of "retrieved information" found relevant in this process is used by the worker in making progress in performance of the knowledge work he/she is involved in.

Thus, here we shall work with three entities of as discussed above, (1) the perceived information needs, (2) the articulated information needs and (3) the information. An *instance of relevance* takes place whenever there is a relation between instances of these entities. The entity "context" is information (Mizzaro 1997), which can play very special roles with respect other information. It is defined as "… any information that can be used to characterize the situation of an entity" Dey (2001, p. 5). The concept of context plays a pivotal role in development of TOKM and shall be discussed later in greater detail.

The consumption of the relevant knowledge leads to creation of new knowledge or update of existing knowledge required for solving the problem-at-hand, in the mind of the knowledge worker. Requirements for utilizing/implementing the solution as well as sharing/communicating the new knowledge to other workers, lead the worker to *articulate* the knowledge and thus *produce* new information. In an organization practicing KM, the new information need to be archived in suitable manner and made available to a larger community of workers engaged in knowledge works of similar or related types.

## The KIF model of knowledge work

The above view of *information usage* in knowledge works is captured in the conceptual model depicted graphically in Figure 3. The model *compartmentalizes* the "performance of knowledge works" into knowledge creation or intellectual activity (the *knowledge creation space* or *K-space*), represented as the apex of a triangle and the support space or S-space providing informational support, in form of systems and artifacts, for the intellectual activities carried out in the K-space. The S-space is further subdivided into *Information space (I-Space)* and *Filter space (F-space)*, forming two apexes at the base of the triangle. Based on these names, for brevity, we shall refer to this model as the KIF model of knowledge work.

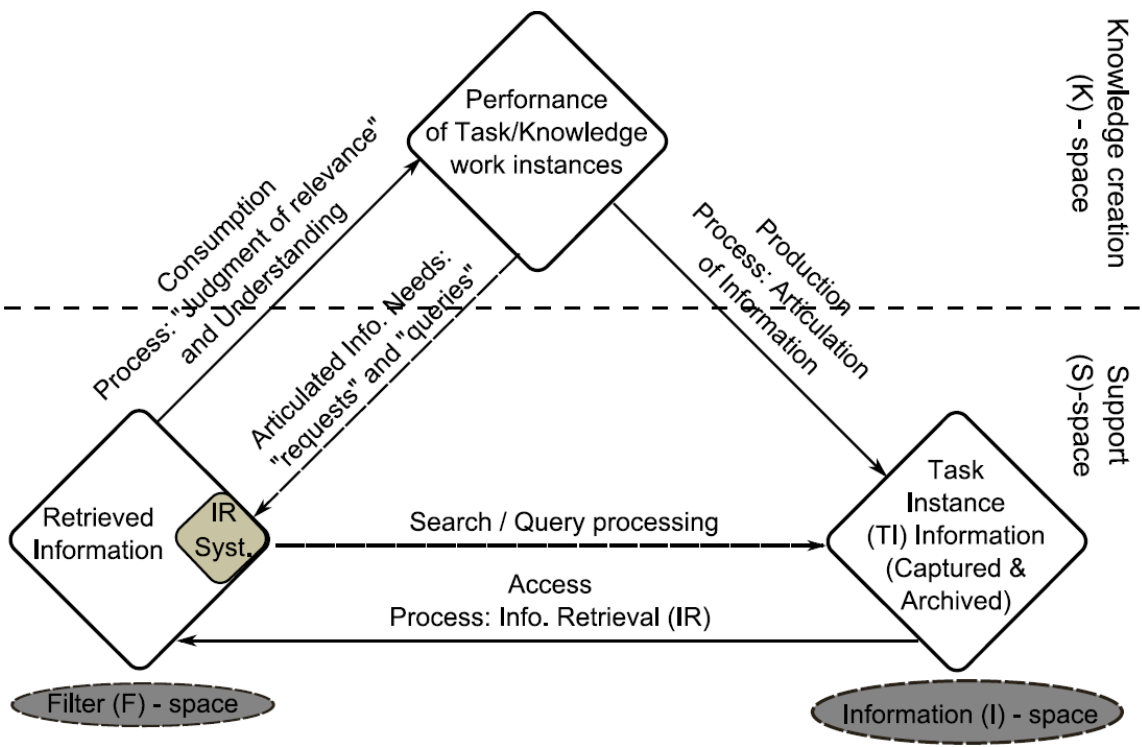

Figure 3: The KIF model of Knowledge work – the information consumption-production duality

## Understanding the "Spaces"

The Knowledge creation space or K-space in short, represents the intellectual activities of human actors engaged in performing knowledge work/task instances. In other words, the K-space represents the *production/reproduction* of "knowledge" (Stehr 1994). They require, in part of the knowledge workers, application of their various cognitive/intellectual faculties, including creativity and reflectivity, for building thorough understanding of the problems and problem environments and finding solutions with the aid of knowledge in their possession as well as new knowledge gathered from the relevant information sought, received and consumed. All these result in creation of new knowledge and/or update of knowledge in the workers' mind. In other words, the processes in K-space work with knowledge to create new knowledge.

The informational supports for the intellectual activities are provided from the support-space or S-space, which deals with information. Of its two components, the information space or I-space represents (1) the facilities for archival of information and (2) the facilities for accommodating efficient search for information by IR systems. Note that, I-space is not limited to only the archives available within the organization, it represents the whole set of archived information available to workers, including external sources such as Internet. Similarly, the filter-space or F-space also has two components, (1) the information retrieval facilities, often constituted of a number of different actual systems compatible with different archives available for access (i.e., query tool for a database, search engine for Internet etc.) and (2) facilities for presentation of retrieved information.

## Understanding the edge processes

Each of the spaces has many internal processes, systemic and/or cognitive, for managing and organizing the entities it represents. As indicated in the KIF model (Figure 3), there are also several inter-space processes. We call these processes collectively the "edge processes" which are of critical importance in knowledge works. The edge processes which connect the K-space with others, involve transformations of knowledge to information and vice-versa. Naturally, they involve significant application of cognitive abilities by the human actors. The edge process representing *production of sharable/communicable information by the acts of articulation* bridges the K-space with I-space. The F-space share two edge processes with K-space. First of them represents *the information needs as articulated by the worker* in order to exploit the IR system capabilities. The second process represents "judgment of relevance" of the retrieved information by the worker in light of his/her *perceived information needs and subsequent consumption of relevant information* found within the retrieved information.

In contrast, the edge processes between I-space and F-space, the *search* and *retrieval* of information, deal with information only. In modern IT-enabled work environment, these two processes possibly enjoy greatest level of system supports. Once the information need is received and processed by the IR system, it conducts the searches of the archives and selects a subset of information, typically embedded within documents, available in the archive for presentation to the worker.

## Available IT support for the spaces and edge processes

It is clear that the K-space is where the human intellectual prowess for *knowledge creation* is at work. Knowledge creation, despite being most crucial part of knowledge works, is not much amenable to information technology support (Polanyi 1967, Nonaka 1994, Markus 2001). In other words, there is not much scope in of using information system *directly* in this core part. They are more susceptible to various social, cultural and behavioral aspects of the organization and the workers. The I-space enjoys large degree of technology support resulting from a long and distinguished tradition of research and practices in areas of large scale archival and indexing technologies. Common technologies include database management, content management, intranet, internet etc. Similarly, in the F-space, the problem of IR is extensively investigated by the researchers in Information Science and host of tools and technologies made available by IT researchers and practitioners. They include a large number of powerful search and/or query tools accessible to the workers. Further, these tools, by themselves or in conjunction with other information presentation/visualization tools (e.g., Web Browsers) do a fairly good job of presenting the retrieved information to the workers. These systems together have made significant capabilities available for information archiving and retrieval.

The edge process of articulation of knowledge towards producing information enjoys some technology support, mainly towards recording of the information in docements. The class of tools used for the purpose is commonly known as "office tools" which includes tools for word processing, spreadsheets, e-mail etc. Of the edge processes connecting K-space and F-space, articulation of information needs is traditionally supported through search and query interfaces of the IR systems. However, more sophisticated interfaces accommodating advanced *relevance-feedback* based approaches (Borlund 2003) or the *belief-revision* based approaches (Lau et al. 2008) are being studied by research community. The other edge process involves identifying relevant information among those retrieved, is a highly subjective cognitive task. However, the way the retrieved information is presented (e.g., graphical, clustered etc.) to the worker often helps to some extent.

## WORKING EFFICIENTLY WITH RELEVANT INFORMATION: THE META-REQUIREMENTS

As noted earlier, the core requirement of the knowledge workers is deceptively simple, *finding and working with relevant information*. Their need of support is having the means to do so *efficiently*. The act of information seeking is dependent on the perceived as well as articulated information needs and the information retrieved by the IR system, which in turn depend on capabilities (e.g., algorithms, indexes used) of the system. Let us examine here the core issues involved in improving the efficiency of the seek process and possible scopes of IT interventions.

To find relevant information, the users need to examine the set of retrieved information to find relevant ones. Obviously, the improvement in this process requires increasing the proportion of relevant information in the retrieved set as well as more efficient means for examining them for relevance. In IR literature the former is know as the "precision" of the retrieval (Saracevic 1999), the ratio of relevant items retrieved to all retrieved items or the probability that a retrieved item is relevant. It is used as a basic performance measure of the IR systems. The other impotant performance measure "recall" is defined as the probability that the system can retrieve a relevant item if it exists in the archive. While good recall is a vital property of IR systems, it can be addressed by the IR system researchers. Here we shall not consider recall (i.e., assume it to be 100%), but concentrate on how precision can be improved from the users' part.

Improved precision can be achieved in two ways, (1) by retrieving information with more proportion of relevant information and (2) by retrieving less total information while containing same number of relevant information. From the users' part, the first way requires correctness in perception and articulation of the information needs. The second way demands crispness or precision in the information needs. However, achieving both of them require that the knowledge worker is in possession of enough knowledge gained through education, training and experience so that he/she can form a good enough perception of the information needs and articulate it effectively to exploit the system support for information retrieval. However, such scenarios, as pointed out by in context of emergent knowledge processes (EKP), more than often do not realize. The potential workers are often "…unpredictable in terms of job roles or prior knowledge" (Markus et al. 2002, p. 179). Thus a support system, to be effective must provide means to overcome the possible lack of experience to a considerable degree.

Searching information for problem solving rarely involve single information-seeking (or simply *seeking*) episode. Typically such activity involves multiple seeking episodes (Belkin and Croft 1992). During these episodes the information-seeking behavior evolves based on the information consumed during previous episodes (Mizzaro 1997). Helping a worker in forming correct perception of information needs result in finding the required information within less number of seeking-episodes and thus improving efficiency. We shall see later how such support can be provided through *contextual information*. The other problem of efficient seeking involves helping the worker correct and faithful articulation of information needs. This problem has motivated studies in Natural Language Interfaces (NLI), those allow machine-processing of natural language "requests", without explicit need in part of the system users to transform the requests into system-language "queries". Frost (2006) provides a survey of the NLI techniques and recent developments in the field of NLI. The systemic aspect of improving precision involves improving the capabilities of the IR systems and is the subject matter of the field of Information Retrieval (IR) and its sub-field Information Filtering (Belkin and Croft 1992, Faloutsos and Oard 1995, Mostafa et al. 1997, Greengrass 2000). However, we shall examine how contextual information can be used for the purpose.

Apart from precision, the other problem faced by the knowledge workers is that of "information overload" (Eppler and Mengis 2004). This occurs with the increase of available information resulting from accumulation over time and/or availability to new information sources. In case of information overload, despite good precision, the user is presented with a large set of information whose judgment of relevance is beyond his/her cognitive capabilities within available time. It has serious implications on the performance of the knowledge workers (Hallowell 2005). There is no obvious solution out of this problem. However, this problem is many-fold aggravated by conventional use of *documents* as *units* of creation, archival and retrieval of information. As we observed earlier, at any given moment, a worker is typically engaged in an activity as part of a larger task. His/her information requirement is dictated by the current activity. In contrast, the information in documents is seldom available at the required granularity level. The worker usually needs to read whole documents and sift through a large mass of information to find some relevant information embedded in small part of the documents. This process puts enormous demand on the time and cognitive capabilities of the workers.

## *The meta-requirements*

Based on the discussions above, here I formulate the meta-requirements for the KWSS and KWSP. While searching for information relevant to the current activity, the user, in role of a consumer of information, needs to spend significant intellectual effort and time in reading and understanding one, or more likely, multiple documents, which contain relevant information in parts (possibly quite small ones) of them. This observation leads us to identify our first meta-requirement:

> MR1: The IT-enabled KWSSes should support access of information at the level of granularity matching the activity-level information needs of knowledge workers.

The information need as *perceived* by a worker, is built upon his/her *perceptions* regarding the problem, the domain, current task and activity. This perceived information needs are formed in the minds of the workers based on their knowledge states. Also, they may be different from the *actual information need*s (in the sense the kind of information can bring about best-possible execution of the activities and consequently the tasks). Of course, the actual information need, in exact sense, is a notional construct. Nevertheless, the closer is the perception to the actualities; the better is the quality of the outcomes of work. Forming such perceptions is largely a matter of experience and other intellectual abilities. However, easy availability of information regarding various aspects of tasks and activities in general as well as on prior task/activity instances of similar and related types can go a long way enabling the worker to easily build a perception of information need fairly close to the actual. Thus, the second meta-requirement can be stated as:

> MR2: A KWSS should facilitate the workers easy access to information regarding various aspects of the tasks and activities as well as historical information regarding how instances of similar and related tasks/activities are performed and what were their effects/results were.

Information is stored in archive(s). To access them, one requires articulating his/her information need. The perceived information need, to be satisfied, needs to be articulated by the worker. The *articulated information needs*, in system-agnostic natural language is known as "request" (Mizzaro 1997) and are best suited for human workers to work with. However, to avail systemic aids for searching information in a archive, the requests must be translated into an unambiguous, formal language understandable by the system. The information needs expressed in system languages are called the "queries". Formulation of proper queries representing the requests, for effective use of the IR systems is one of the significant challenges faced by workers. This leads us to formulating the third meta-requirement:

> MR3: A KWSS must facilitate easy conversions of natural language "requests" into system language "queries".

Typically there are many types of the "surrogate" information related to the documents, e.g., title, author name, keywords etc. are organized to form easily searchable indexes, catalogues etc. to provide fast systematic search capabilities. As we discussed earlier as well as formulated in form of MR1, the knowledge workers, while seeking information, can work better with information at a level of granularity typically not entertained in documents. So if we need to search the information at proper granularity level, system support must be available in form of surrogates and their indexing at the same level. Thus, our fourth meta-requirement is:

> MR4: In a KWSS, the system must support working with surrogates of information at requisite granularity level.

### Producing relevant information

Production of information is essentially articulation of the knowledge gained by performing the knowledge works. The objective of producing information is to make them available for consumption. The produced information to be useful, a worker as the producer strives for *accuracy, objectivity, authenticity* and *display of rationality of thought* (Schultze 2000) in the product. From linguistic perspective, information is considered *rhetoric*, which according to Webster dictionary is "the art of speaking or writing effectively" and/or *discourse*, which again literally means "formal and orderly and usually extended expression of thought on a subject (Webster dictionary)". Clearly, producing information is a significant intellectual activity requiring high level of relevant skill set. This leads us to formulate the fifth meta-requirement as:

> MR5: A KWSS must support workers in articulating their knowledge to produce information at proper granularity level with accuracy, objectivity, authenticity and display of rationality of thought.

### *The Platform-level meta-requirements*

The five meta-requirements deduced above are at the level of individual KWSSes. However, the objective of the TOKM is building IT platforms, which can allow building, maintenance, modification and interoperation of multiple KWSSes on top of them. Thus our platform-level meta-requirements can be formulated as follows:

> MR6: The KWSPs must facilitate building, maintenance, modification and interoperation of multiple KWSSes built on top of them in such a way that:
> MR7: Each of the individual KWSSes will be able to satisfy the meta-requirements 1-5.

# PUTTING TOGETHER THE ELEMENTS OF THE META-DESIGN

A meta-design, according to Walls et al. (1992) is consisted of the *artifacts* which can be *hypothesized* to meet the meta-requirements. For example, in case of Relational Database Theory (Codd 1970) the meta-design is "… a set of tables in third (or higher) normal form" (Walls et al.

1992, p. 43). In the following we shall develop the appropriate artifacts to satisfy the 7 meta-requirements developed above.

### *The comprehensive solution: richly contextualized information*

The approach to solution proposed in the current paper is based on extensive use of *richly contextualized information*. In the following we shall first try to understand the concept of "context" from the informational perspective. Then we shall proceed to examine how the contextualization of information and its proper use can be achieved so that the set of meta-requirements developed above can be satisfied.

### Context is information about "information"

Context is a fairly complex concept with many facets (Dourish 2004). However, much of it is commonly regarded as information (Mizzaro 1997), and thus codifiable/recordable. Context, as information, render greater meaningfulness and usability to other informational entities (Dey 2001). With recent advent of mobile and embedded systems, where to be effective, the user as well as the systems need to be aware of the "situational" and "environmental" contexts, has attracted the attention of research community towards the class of system known as *context-aware systems* (Schilit and Theimer 1994, Dey et al. 2001).

From the perspective of information usage in knowledge works, the production of information has situational, environmental and other contexts which the knowledge worker, as producer of information, finds himself/herself immersed within. On the other hand, a worker, as a consumer of information, has his/her own context, which shapes the information seeking behavior. Naturally, to make proper use of the information retrieved, the consumer need to know the context of production of the information, compare and understand them in light of his/her context of information needs. To enable such an environment where the contexts of production and consumption of information can be easily *identified, compared* and *understood*, the system must support the producer to efficiently produce information along with context of the production. Further, from consumer's perspective, the system must support easily identifying the context driving the information seeking and compare it with that of the retrieved information.

### Contextualization of Information

The information accompanied by its context is known as "contextualized information". Let us call the act of "producing information along with context" as the "contextualization of information" or "contextualization" for brevity. Now, let us examine the issue of contextualization with respect to knowledge works. We can readily consider two distinct levels of contextualization of information with respect to knowledge works. First of them is the contextualization of information at the *categorical level* of tasks and activities. The notion of categories is central to our cognition (Harnad 2005). Here the category contexts allow users to *describe* as well as *understand* for a given piece of information, the characteristics of activity and task types within a broader perspective, whose performance has resulted in production of the information. The second level of contextualization occurs at the *instance level* of tasks and activities. Here the contexts refer to the *relationships* of a piece of information with other pieces of information and activities at level of instances of performance.

## *The meta-design artifacts*

For enabling system support for contextualization we consider the classes of artifacts listed in Table 2.

- **MDA1:** The KWSS-specific guidance/description artifacts.
- **MDA2:** The connectivity or association artifacts.
- **MDA3:** Creation, archival and retrieval of information, which are linked, supported and provenanced at required granularity level.
- **MDA4:** An environment for efficient accommodation and use of the above artifacts.

Table 2: The meta-design artifacts

## The guidance artifacts

The guidance/description artifacts for a particular KWSS are the resources to describe the target task/task-type. They provide support in two ways, (1) allowing a novice worker (i.e., someone not very familiar with the *particular* task-type supported by the KWSS; he/she may be proficient with other task-type(s)) to quickly understand and learn about the task and (2) guiding a worker to during the performance of the task. Each guidance artifact is composed of smaller components supporting categorization of various entities from several perspectives. *Categorization* or *classification* is one of the most crucial aspects of intelligent behavior (Bowker and Star 2000). It pervades every sphere of human endeavor. The guidance artifacts, enabling categorization of various entities as well as their relationships allow the knowledge workers to understand from multiple perspectives the typical nature of the information as well as activities and knowledge works responsible for producing them. In this paper, we shall consider three guidance perspectives, the *task and activity,* the *informational dependency* and the *domain vocabulary* perspectives as of prime importance.

The task and activity perspective of a knowledge work allows the workers to learn about the task performance in an organized manner as a *system of activities* (Burstein and Linger 2003). On the other hand, the use of artifacts describing the informational dependencies allow the worker to study how various pieces of information are developed and utilized at activity level towards accomplishing the larger goal of the task instance. Lastly, the availability of domain vocabulary as supporting artifacts, allows a user to understand the semantics of the elements of knowledge works. For example, consider the guidance artifacts for a task of patient-care. When a doctor starts treating a patient, first he/she performs the activity of *examination* which leads to production of informational element *the result(s) of examination*, which, in turn, is known in the medical domain vocabulary of the health professionals as the *first impression*. Clearly, these three entities are components of there different perspectives of the knowledge work and their nominal characteristics can be captured in the respective guidance artifacts.

## The association/connectivity artifacts

The association or connectivity among pieces of information is a powerful idea. Extensive use of *hypertext* documents carrying navigational links embedded within them bear witness to the fact. In TOKM, the association artifacts establish the correspondences or linkages of different types among various entity-types. At the task instance (TI) level, (a) the associations among instance-specific informational elements describe their interdependency as well as contributions in achieving the goals of the task and activities performed; (b) connections among the artifact elements across guidance perspectives, establish correspondence among the guidance elements enabling the potential users to examine informational elements from different perspectives; (c) the association between components of a guidance artifact for a particular perspective caters for understanding the nature of the task from that perspective; and (d) the links between instance-specific informational elements and guidance artifact elements provide means to identify the informational elements with well-defined categories.

Continuing with the patient-care task example, from activity guidance perspective, *examination* is followed by, i.e., linked with, *determination of possible diseases*. The same from informational dependency perspective, the *results of examination* forms the basis of forming the *list of possible diseases*. Again linkage between elements across perspectives, say, *examination* (an activity) and *results of examination* (an informational element) establish their correspondence. At the instance level, the instance-specific informational elements such as *high temperature, headache etc.* found in examination will be associated with the activity *examination* and informational element type *results of examination* as well as with instance-specific information on possible disease *influenza*; which, in turn will be associated with guidance elements *determination of possible diseases* and *list of possible diseases*.

## Creation, archival and retrieval of granular information

As pointed out earlier, *documents* are the units of archival as well as retrieval in conventional information management systems. However, at the activity level, the information requirements of the workers are more focused, which are difficult to serve efficiently with document-centric systems, especially under constraints regarding availability of time and other resources. This problem is further aggravated with possibility of information overload. A plausible solution out of this lies in coming out of the idea of documents as the *only possible unit of creation, archival and retrieval* of information. We must consider using units of information archival and retrieval, more compatible with the activity level demands. Let us call such units the *informational elements* (IE) (Schultze 2000).

In the production side also articulation of IEs as they are developed during activities, relieve the knowledge worker from the significant effort and time to be spent in putting together the IEs in form of a document. However, IEs, when put together in a document, become contextualized to some extent. By dealing with them individually the worker faces the prospect losing it. This problem with informational practice involving granular information was observed by Ruhleder (1994). Nevertheless, the extensive contextualization of IEs as envisaged here (1) by association with the components of related guidance artifacts and more importantly, (2) by association at the task instance level among themselves in order to accomplish rich contextualization in terms of three vital attributes of information, *proper granularity, extensive support* and *verifiable provenance*, do not only compensate for the loss but also introduce additional richness of context to work with.

## Creating an environment for accommodating and exploiting the artifacts

To develop a suitable environment to accommodate the above artifacts, let us consider as our final artifact, *an extension of the KIF model of knowledge work* as depicted in Figure 4. We shall call the new model *eXtended KIF or XKIF model of information usage* in knowledge works.

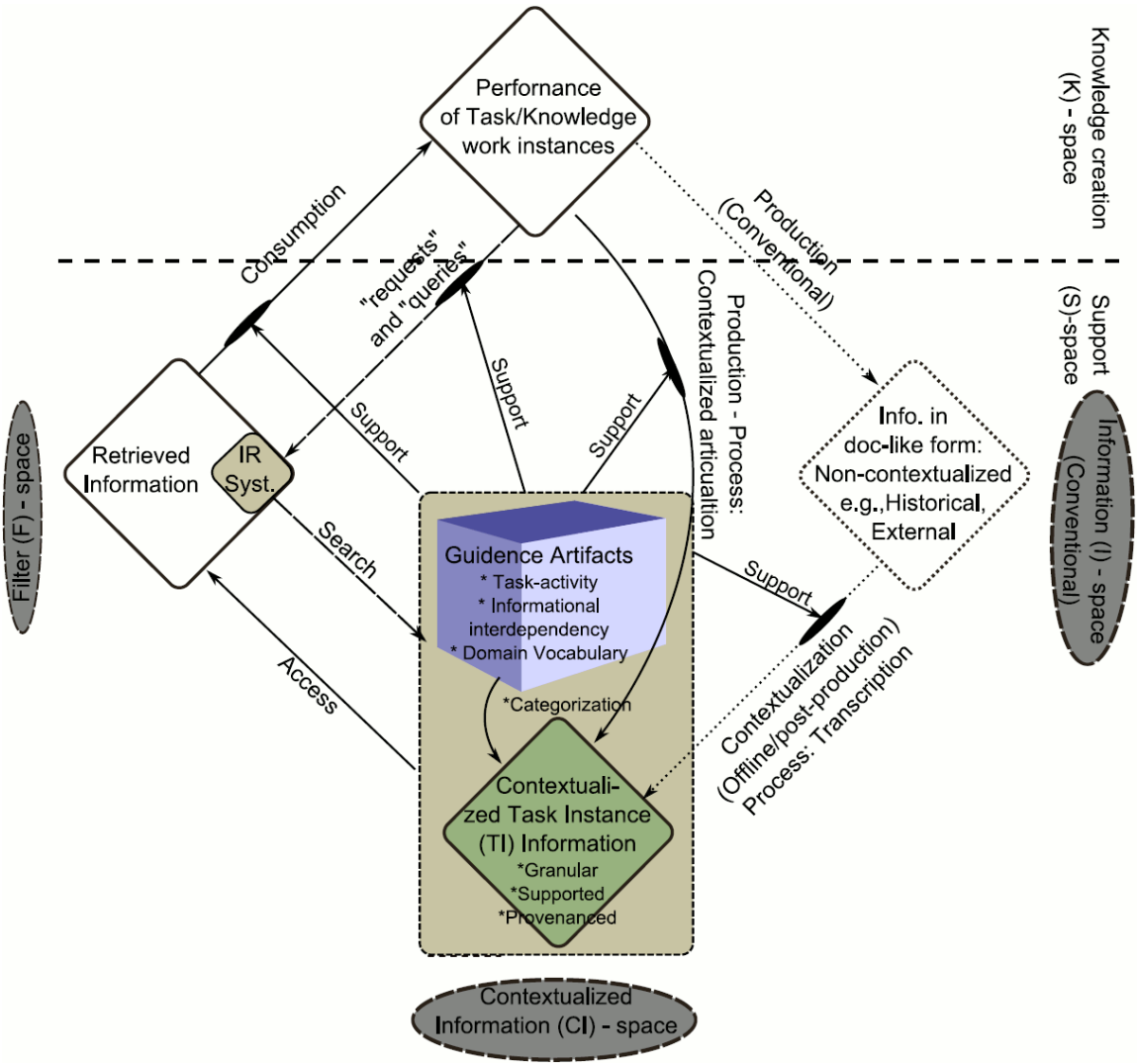

Figure 4: The eXtended KIF (XKIF) model for *contextualized information* usage in knowledge works.

The most significant aspect of the XKIF model is the replacement of the information space or I-space, with the *contextualized information space* or CI-space for the task types for which KWSSes are implemented and in use. However, the conventional I-space also co-exists to represent historical information, information from tasks not yet supported by KWSSes and information from external sources. Naturally, such information has document-like forms and is also required while performing the supported tasks. Thus, in the XKIF model I introduce a process *transcription* for moving required information from I-space to CI-space. The process is described later.

The CI-space represents an archive for information containing guidance artifacts for one or more tasks/task types as well as the information generated by performance of the instances of those tasks. The CI-space also maintains the association or connectivity artifacts used for establishing various types of connectivities. Thus, the task instance information in CI-space is in granular and contextualized form associated with components of respective guidance artifacts as well as interconnected among the informational elements to achieve contextualization in terms of support and provenance. In the next section we shall examine how the design artifacts developed above address the meta-requirements.

# TESTING THE DESIGN ARTIFACTS

Walls et al. (1992) indentify the component of the design theory *testable design product hypotheses* as a set of hypothesis-level tests for determining whether the meta-design satisfies the meta-requirements. Such tests can take many forms. While in some of the ISDTs, e.g., Relational Database Theory (Codd 1970), the design artifacts can be tested analytically, not all of them can have analytical formulation leading to such mathematical/logical tests. In case of TOKM, I demonstrate the capability of the artifacts in addressing the meta-requirements, using the method of *descriptive and informed arguments*, identified by Hevner et al. (2004) as one of the alternative evaluation methods often used in design theory research.

## *Addressing the meta-requirements*

The information contained CI-space, due to their organization as described above, can easily be leveraged to provide significant supports to several processes, especially edge processes connecting the K-space, which are difficult to support in conventional way. In the following I shall examine how various kinds of utilization of the CI-space will support the processes and thereby fulfill the meta-requirements listed in the previous section.

At the platform level, each KWSS for supporting a particular task type can be enabled by creating and deploying in the CI-space the set of guidance artifacts for the target task type. Further, the platform also provides means to access the guidance artifacts as well as task instance information and information on various interconnections among them. These facilities together satisfy the meta-requirements 6 and 7.

Maintenance of the CI-space requires usable means for creating information in a form consistent with its organization. In other words, we need to facilitate articulation of knowledge by the knowledge workers in granular and contextualized form. The process is depicted as the "edge process" *contextualized articulation* in Figure 4, as an alternative to the conventional production of information in the KIF model of knowledge work. We can easily see that since the knowledge workers perform at the activity level, production of granular information compatible with the activity-level knowledge creation is actually easier than creating information in form documents by consolidating the information from multiple, possibly large, number of activities. However, to contextualize the granular information to reflect the objectivity and authenticity, and to do so efficiently, a worker needs support in indicating various supporting information as well as provenance information. In this case, such support can be easily made available from the CI-space. The support can include (1) categorization of the produced IEs by associating them with the relevant components of the guidance artifacts available in CI-space for the task type; (2) identifying IEs in CI-space supportive to the new IE by associating it with the existing IEs found relevant developing the knowledge whose articulation led to the production of the new IE; and (3) while producing an IE, recording along with other contents of the IE, the existing situational information to provide valuable provenance information. However, additional provenance information also gets recorded implicitly via various associations/links. These demonstrate the capability of addressing the MR 5.

Also, in real world, we can not expect all the information to be available in the CI-space automatically. The above support is available only to the workers working on the task-types for which KWSSes are implemented and only since the time they were made available to the workers. However, a worker may need information developed earlier or through performance of unsupported tasks (they include external information). This information is represented by conventional I-space. To make effective use of such information in the new environment, they need to be transformed to be compatible with the CI-space. This can be achieved by means of the process *Transcription* shown in Figure 4. This is an offline or post-production contextualization of non-contextualized information in typically document-like forms. As indicated in Figure 4, transcription can also be efficiently supported by the contents of CI-space in ways pretty similar to those described above for supporting contextualized articulation, which can also be called the online contextualization or contextualization-at-source process, in contrast to the transcription process.

Once the information is made available in granular level, developing the typical surrogates is largely an engineering problem not very much different from those used at conventional document level. However, processes of creating contextualized information (i.e., the contextualized articulation and transcription) described above automatically perform multiple categorizations of the IEs with the help of the guidance artifacts. Such rich categorization can serve as very effective surrogates for the IEs in (1) helping human workers to conduct quick preliminary judgment of their relevance and (2) the systems in implementing highly precise search methods. Thus, the meta-requirement 4 can be met easily.

The capability of knowledge workers in articulating the information needs, which will ultimately drive the search algorithms in the IR systems, is crucial one. However, in general, effective mastering of this capability is difficult. It involves first understanding the needs in an objective manner, which the worker can articulate in natural language. To use a typical IR system, the worker needs to translate it into system-acceptable query language. Several advanced IR systems provide interfaces to express the needs in natural language itself. However, even the natural language vocabulary acceptable to a system can not be expected to conform to the vocabulary of *all possible users* of the system. To address such difficulties, as envisaged in meta-requirement 3, interfaces can be designed to incorporate many functionalities for hint-generation, recommendation etc. supported by the CI-space contents, to help the user expressing the information needs in system vocabulary.

The model supports very effective and deep learning about the supported tasks (i.e., for which individual KWSSes are implemented and in use) as demanded by MR2. Firstly, access to the relevant guidance artifacts allow the workers to learn about structure and processes of performing the tasks. Secondly and more importantly, task instance information, the way it is organized in CI-space, on analysis can lead to deeper knowledge about the strengths and weaknesses of the practices in performing the tasks. Proper utilization of the design in analysis can create opportunities for *double loop learning* (Argyris and Schön 1978), a very crucial but extremely difficult to achieve requirement for all organizations.

Lastly, due to the existence of the CI-space, access to the granular information results naturally. The IR systems can be configured easily to enable them to access granular information and thus to satisfy the meta-requirement 1. Actually, given the level of support available in the CI-space, one can very reasonably expect that with little effort the information retrieval from the CI-space can be made much more efficient than retrieval from traditional I-space.

# REALIZING THE ARTIFACTS WITH TECHNOLOGY: THE DESIGN METHOD

In the *design method* component of TOKM, I outline an IT platform, the "Knowledge Work Support Platform" or KWSP, which can realize the design artifacts for building *real systems* which can meet the meta-requirements. The KWSP is presented here in form of a generic architecture as depicted in Figure 5. It has a layered architecture, where the each layer brings in new components and capabilities for leveraging the functionalities of the inner layer(s) and thus, from human user perspective, value-add to those in the inner or contained layers. The outermost layer represents the KWSSes which the knowledge workers can use for performing their tasks. In the following I shall examine the technological realizability issues of the architecture drawing upon state-of-the art of IT research and practices. Along the way, I shall also point out various relevant research problems. We shall start out discussion from the innermost layer and move outwards from there. Note that, the discussion here addresses the issues in conceptual and to some extent logical levels. We do not venture into the particularities of the IT tools and technologies to use. For example, we provide no recommendations here on whether a relational database tool or a XML document repository should be used for archival. Such are engineering issues and we leave them at that.

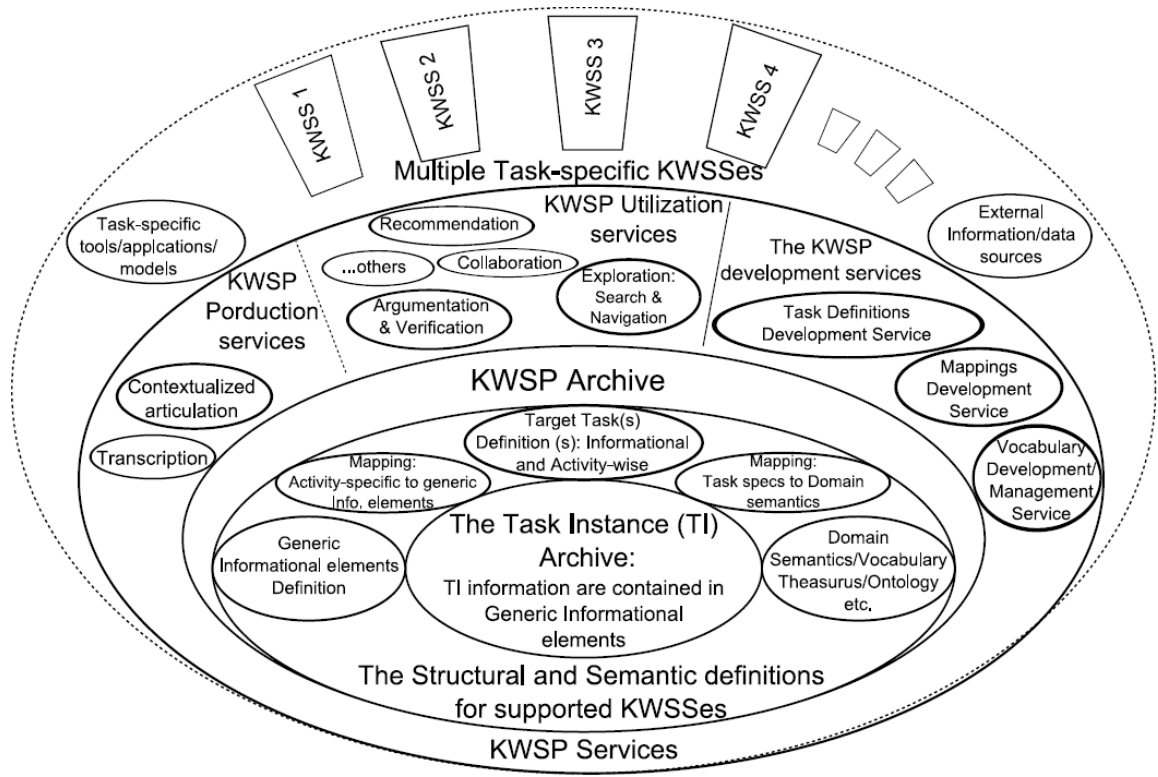

Figure 5: The layered architecture of KWSP components and KWSSes

## *The Task Instance Archive*

The innermost layer in the architecture is the *Task Instance* (TI) archive that stores the informational elements developed during the performance of various instances of knowledge works. I propose that each IE be stored in a *container object* of a type belonging to a *class of generic information containers*. A hierarchy of generic information container classes can be developed from the consideration of generic IE categories developed at activity level. This will allow us to re-use the container types across the task and activity types as well as instances. The basis of such an approach is discussed below.

## Generic categorization of tasks, activities and informational elements for re-use

Burstein and Linger (2003) categorized their TbKM case studies along the attributes (1) application area, (2) task and (3) issues investigated. Their categorization can be generalized by using little different (but related) attributes, (1) *generic task types* such as *decision making, research, planning, strategizing* etc.; (2) *application area* or *application domain*, e.g., *health-care, finance, telecommunication* etc.; and (3) *tangible outcomes* of the performance of a task instance. For example, consider a patient-care KWSS that supports a doctor to treat patients. The generic type and application area of the task are "decision making" and "health care" respectively, while the tangible outcome (at least the major one) of performance of one instance, i.e., treating one ailing patient, is to get him/her in a state of health. Note that, the categories are not absolute and may somewhat overlap. Continuing with our example, while treating a patient the doctor might need to carry out some research as well as the financial planners of the hospital may want to know the expenditures and payment patterns, which has dependence upon how the treatment has been carried out.

Similarly, as discussed earlier, at the activity level also we can think of categorization of the informational elements corresponding to generic types of tasks and activities. We present some of the examples of generic IE containers below:

- Observation: representing facts, e.g., measured entities, their semantic, methods and sources.
- Findings: information based on aggregation of correlated facts.
- Analysis: an understanding of problem based on one or more findings.
- Hypothesis: a tentative estimate of problem status in an informed way.
- Decision: a solution based on the verification of hypotheses and other IEs.
- Plan: for implementation and/or follow-up of the implementation of a decision.

Such categories can be studied in order to design suitable IT support for working with them in a re-usable manner. Even, we can think of developing high level template-like designs for IT support of generic task types using the above generic IE classes which can be re-used (with suitable customization) across application domains. The specifications of application domains bring with them the vocabulary of the domain. For example, in the patient care case, a hypothesis can represent a possible ailment of a patient, which the domain vocabulary designates as the *possible condition*. Similarly, a diagnosis is a decision while the treatment and follow-up plans can be modeled by the generic IE type *plan* or with its specialized subtypes.

The approach of modeling the granular level IEs as contents of generic information container classes has several advantages. Following Object-oriented design practices, we can consider a set of re-usable types for container objects which can accommodate various general as well as category-specific characteristics of their contents in an organized manner. Further, the containers as objects can easily accommodate references or links to other containers to implement various relationships among the IEs. Given the proper design, the implementations can be fairly straight-forward.

## *The "Definitions" layer*

The definition layer is part of the KWSP archive and stores the KWSS-specific definitions. They include various structural and semantic definitions of the target task types and their mappings to the generic container types to enable re-use of the generic IE containers for all the supported KWSSes in transparent manner. These definitions and their proper utilization are at the heart of the KWSP. There are three types of interconnected definitions, (1) the *task-activity structure*, (2) the *informational relation structure* and (3) the *domain vocabulary*. These are extremely important in helping the knowledge workers in their performance of tasks and activities as well as allowing them to learn about unfamiliar tasks and activities (Hartley 1998).

The tasks or knowledge works are known to be very difficult, if not utterly impossible, to *formally* structure in terms of logical and/or mathematical operations (Simon 1973, Keen and Scott Morton 1978, Markus et al. 2002), preventing their automation. Nevertheless, tasks do have structures, albeit informal, imprecise and often incomplete to a varying degree (Hartley 1998). Here our aim is to "informate" rather than to "automate" (Zuboff 1988). Thus, ready availability of such structural information about the tasks and activities can be of immense value. In TbKM, the tasks are defined as systems of activities and decomposed in structures and processes at activity level (Burstein and Linger 2003, Aarons wt al. 2005). Similar approach can be followed for building the activity-level structural definitions of the knowledge works. Further, task-structures, known as *task analysis* (TA), are studied in the field Human-computer Interaction (HCI) also for designing *interactive IT systems* (Card et al. 1983, Payne and Green 1986, Johnson and Johnson 1991). The *Activity Theory* based TA approach in HCI (Nardi 1996) accommodates human actors as integral components or "protagonists" in the system environment. It shows good promise in analyzing complex knowledge work scenarios for building the definitions as well as in designing easy-to-use system interfaces for performing such works.

However, it must be remembered that the knowledge of the nominal structures of tasks, though very essential for the human workers for performing them, their roles are more of *guiding rather that binding*. It is often found that in order to perform a task instance, the worker needs to deviate from the nominal structure led by creative thinking and deliberation/reflection (Johnson and Carruthers 2006). Such deviations, rather than being violation or reflective of inadequacy of the nominal structures, are of necessity. The necessity arises from the fact that given continuous and sometimes quite rapid changes in social, economic and technological environments, while performing a task-instance the worker may face some novel challenges. He/she needs to apply *ingenuity* to face the new challenges leading to adaptation to the new scenario. Vicente (2000) discussed this issue in detail with the help of an interesting case study. The ability to adaptation sometimes referred as the "learning to learn" is of crucial importance for the workers individually as well as from the organizational viewpoint. This poses a problem in using the conventional workflow technologies for building and deploying these structures (Dustdar 2005).

Nevertheless, even in the face of significant deviations, the knowledge of the nominal structure works as a *frame of reference*. In this regard, the roles of the nominal structures of tasks and the deviations can be easily related to the influential "Theories of action" (Argyris and Schön 1978) for organizational learning. We can easily recognize the nominal structures as the *espoused theories of action*, while the information on how the task instances are actually performed; including deviations, are the *theories in-use*. A careful study of them can be a powerful way of achieving *double-loop learning* in the organization.

Information, being product of knowledge works; it is easy to understand that their relationship structures are related to the activities where the pieces of information are produced. The relationships among the IEs reveal various roles played by the IEs with respect to one-another. The notion of structures in *Information* or *articulated knowledge*, especially in natural language textual forms can be derived from theories such as "Rhetorical Structure Theory" (RST) (Mann and Thompson1988) and "Theory of Discourse" (Grosz and Sidner 1986). These theories form foundations for studies in various *Computational Linguistics* (CL) problems, including text analysis, text generation, cross-language translation etc.

There may be several types of relationships among IEs (in RST more than 40 types of relations are considered), but here we consider two types of relations among IEs. Firstly, one IE might have its genesis in order to satisfy, (may be partially) the need of developing another IE. We shall call it a *demand-satisfaction* (DS) relationship. The other one is that of support; where an IE may not be created explicitly to fulfill the need of other, but provide referential support for the development of the later. Let us call it *reference-support* (RS) relationship. These two relations are modeled following two important structural relations identified in theory of discourse, the *dominance* and *satisfaction-precedence*. Since the theory deals at the level of *intentions* (behind the production of information), dominance is defined as "… an action that satisfies one intension, say DSP1 (DSP: discourse segment purpose), may be intended to provide part of the satisfaction of another, say DSP2"; similarly, "… DSP1 satisfaction-precedes DSP2 whenever DSP1 must be satisfied before DSP2" (Grosz and Sidner 1986, p. 179).

Enabling the IT system users to work through domain vocabularies which they are much more comfortable with, compared to system-specific vocabularies, is a major area of investigation today. Much of the works in this area is directly or indirectly associated with the problems of realizing "Semantic Web" (Berners-Lee et al. 2001) initiative led by World Wide Web Consortium (W3C). Nonetheless, many of the concepts, understandings and technologies developed by these works can be applied to the area of enterprise computing. Actually, due to the relatively controlled environment of enterprise systems, many of those techniques can work better compared to web, where they have to deal with a hugely distributed and amorphous environment.

There are several ways of capturing and utilizing the domain vocabulary. While artifacts such as dictionaries and thesauri are often used for encoding the domain vocabulary and presenting them to the user, in the last decade the idea of using *ontology* (Gruber 2008) for the purpose has gained a lot of ground. Use of ontology for creation and management of domain vocabulary can give a very versatile way of making the system vary user friendly. However, substantial effort is required for building a good ontology. Further, much of the technologies developed for building and deploying ontology, lean towards accommodating interoperability and understanding of semantics of information at the machine level. However, they can be easily adapted to provide support to human workers in dealing with information from domain vocabulary perspective. The problem is often studied in information science in context of creation and understanding of "annotations" by human actors. A review of works in the area as well as a formal model of annotation can be found in (Agosti and Ferro 2007).

The architecture provides support for development of KWSS through the KWSP development functionalities/services. They can be used for building and deploying the definitions and mappings for implementing new KWSSes as well as modifying existing KWSSes. From the above discussion, it can be easily seen that there exists a body of works to draw upon for creating and deploying nominal definitions of the target tasks of sufficiently complex natures in terms of activities and information structures, supported by well organized and accessible domain vocabulary. A graphic representation of high level structures in the patient-care task is depicted in Figure 6.

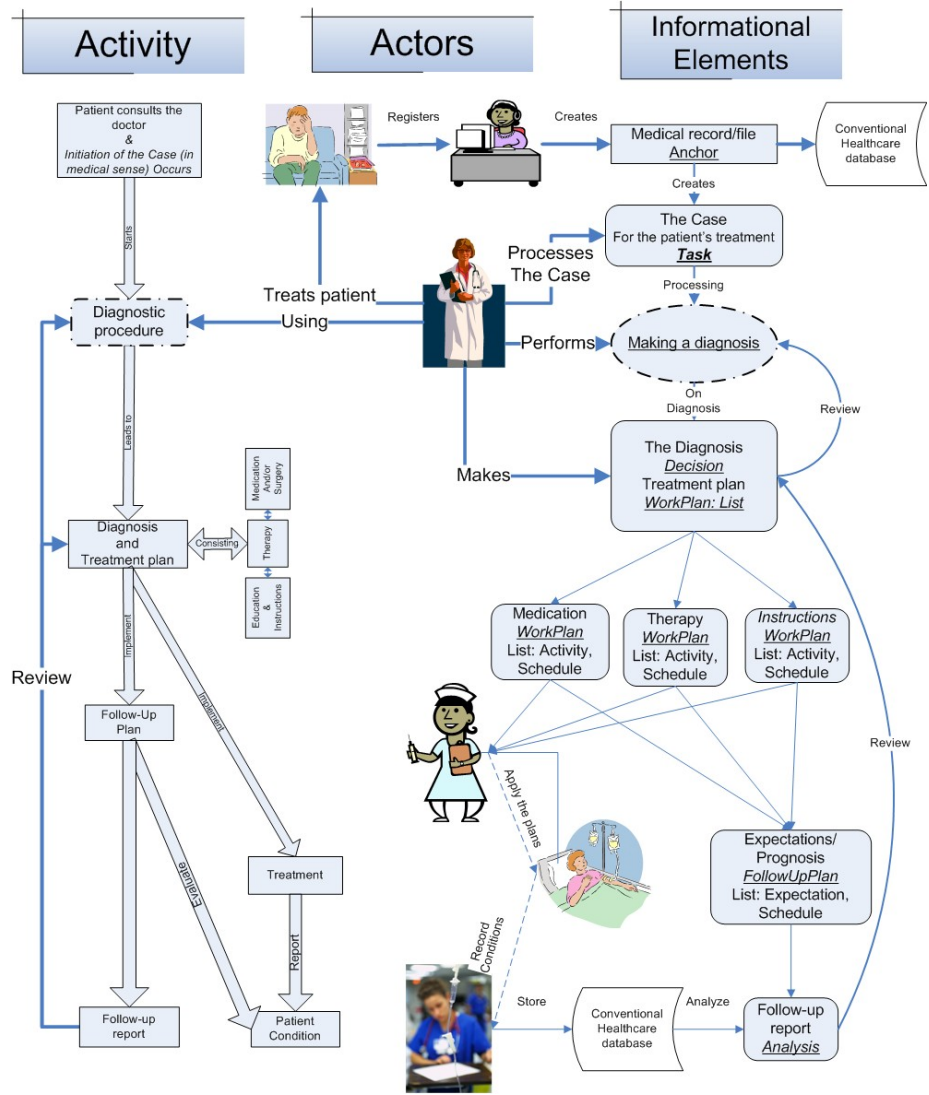

Figure 6: The high-level activity and informational structures for the patient-care task

## *The "service" layer*

The KWSP archive provides a centralized archive of contextualized information and the structural and semantic task-definitions for the KWSSes. We assume that each user or knowledge worker, involved in performing a task of one of the supported types, works with some "Workspace-like" interface, through which he/she can access various kind of information at definition level and task instance level. Let us examine some of the supports those can be delivered to the workers by exploiting the resources available in the archive.

## The KWSS development services

The KWSS development services provide interfaces to the KWSS developer to create, deploy and maintain the task-type specific definitions, vocabulary and mappings for implementing individual KWSSes. Creating and deploying structural definitions of transactional and or operational tasks has been studied and implemented for quite some time using "Workflow Modeling" (vanderAalst and vanHee 2004) techniques for business process management (BPM). However, they demand stringent adherence from the user, which, as discussed earlier, not suitable for knowledge workers. Thus, we need to work on introducing capabilities of systematically accommodating deviations in workflow modeling techniques for using them in KWSS development services (Dustdar 2005).

Creation and maintenance of vocabulary in form of taxonomies such as ontology is an active area of research and development (Corsar and Sleeman 2007). There are several tools available in this area. Prominent among them are the Protégé, developed by researchers in Stanford University and the Unstructured Information Management Architecture (UIMA), developed by IBM Corp. Both of the tools are freely available and widely used by the research community as well as the IT practitioners. However, they are oriented towards dealing with vocabulary artifacts mainly for consumptions of the machines. However, in case of KWSP/KWSS, they need to be comprehendible by human workers as well as usable by the machines. Thus there is a scope of research here in bringing together the above techniques with those deals with creation and management of human-comprehensible "annotations" of information (Agosti and Ferro 2007).

## The "Workspace" services

The workspace services manage the "workspace interface" or simply the "workspace" which allows the workers to interact with the system. The workspace, once the worker identifies the task type and instance he/she is working with, can be configured on the basis of the available structural and domain vocabulary definitions of the particular task type, to support systematic performance of the task instance(s). This facility can guide the worker in his/her progress through the task from activity to activity as per the definitions. As a result, the workspace will be able to maintain the information on "situational context" of the worker with respect to the task and activities. The workspace allows the workers to invoke other services from it. Thus the context information can easily be exported to the services to enable them to work in more focused manner.

The basic facilitation of the above in an IT-based interface is a fairly straight-forward engineering problem. However, there are a lot of opportunities for introducing more sophistication in the design which can help the users to exercise their cognitive faculties. Fortunately, as noted earlier, in recent time the issue of designing system interfaces accommodating cognitive aspects of work has been taken up earnestly by the HCI research community. In our view, possibly most promising developments can be expected from the HCI researches based on the *activity theory* (Nardi 1996), which attempts to accommodate human contributions in the working of the information systems.

### The "Exploration" services

The exploration services facilitate the crucial requirement of search and navigation for information in the archive. The information retrieval (IR) techniques are of main interest here. Though, traditionally the IR systems are document-oriented, at the techniques and/or algorithms levels, the most of the IR problems are formulated as various operations in suitable multi-dimensional vector space (van Rijsbergen 2004) based representations of the archived information. Since such formulations do not depend, at least explicitly, on any particular *unit* of information, they can be easily modified to work with information in granular forms other than document.

Further, due to the richness of the organization of information in the KWSP archive, there is very good possibility that advanced IR features such as *semantic search* (Mangold 2007), *latent semantic analysis* (LSA) and *latent semantic indexing* (LSI) (Landauer 2007) etc. can be used efficiently. There are also possibilities of developing using other natural language processing (NLP) based techniques for matching the queries and contents semantically (Faloutsos and Oard 1995). Further, in recent time techniques of network analysis (Brandes and Erlebach 2005), including those of social network analysis (Carrington et al. 2005) are being developed and used extensively to unearth interesting results from large scale networks such as Internet. The extensive linkage among the elements of the archive opens up the opportunities for analyzing the contents of the archive and using such techniques separately as well as part of the exploration services.

### The "argumentation" service

Argumentation reveals the *reasoning process* behind arriving at a point of view. Naturally, this is of utmost importance in context of knowledge works. In KWSP the argumentation service facilitates (1) demonstrating and recording reasoning processes along with supportive evidences and (2) verification of arguments by following the reasoning processes and if required, independently validating the supporting evidences.

In recent time "argumentation" has become a much studied topic in the field of artificial intelligence (AI) (Bench-Capon and Dunn 2007) in context of autonomous agents. However, here the argumentation service is aimed at use by human actors. Thus, in KWSP, possible approaches for building the argumentation service need to be different from those in AI. One of the possible approaches can be based on formalisms such as Issue-based Information Systems (IBIS) (Kunz and Rittel 1970, Rittel 1980) or gIBIS (graphical IBIS) (Conklin and Begeman 1988).

### The "recommendation" service

People make extensive use of myriad types of "recommendations" in their everyday life whenever faced with making choices without sufficient knowledge/experience to evaluate the alternatives. Knowledge workers also often face such situation while performing their works. Naturally, availability of a system service to present recommendations in form of hints and warnings about various aspects of their work can be extremely valuable for the workers. In KWSSes the recommendation service can be used in various ways, including indicating completeness status of the activities, possible next steps, suggestion about the information needs and many others. An effective recommendation service can significantly lower the expertise/experience threshold required by the workers to perform at desired level of efficiency.

From technology perspective, many of the required functionalities of the recommender service are studied in context of "recommender systems" (Resnick and Varian 1997) or "collaborative filtering systems" (Goldberg et al. 1992). Typically, the recommender systems are used in E-commerce applications for helping prospective customers to make choices. In case of KWSS the situation is somewhat different from such applications. However, many of the differences can actually impact favorably. Most important of them is the possibility of the system to leverage the context information from both sides (i.e., requirement of the worker via the situational context maintained by the workspace and the informational contexts maintained in the archive) to provide high quality focused recommendations. Recently, Adomavicius et al. (2005) studied the use of contextual information in recommender systems.

## The "contextualized articulation" or "articulation" service

With the availability of the above services, the support for articulation of the knowledge to produce information and contextualization of the produced information can be realized in a fairly easy manner. Consider the scenario where a knowledge worker produces one *informational element* while using a KWSS.

The worker works through the workspace interface, which is driven by the definitions for the task-type and keeps track of the *progress* and *instantaneous situational context* of the work. Thus, it can be a simple matter to design the articulation service to access the context information from the workspace and use them to associate the produced IE with proper elements of the definitional artifacts as well as other IEs produced earlier within the task instance. Similarly, the other IEs acting as the supportive/referential evidence for the newly produced IE are accesses by "exploration" of the archive. Thus the design of the articulation service can easily accommodate association/linkage with them by facilitating marking and importing the relevant IEs to the workspace and creating the associations. For producing complex IEs, the "argumentation" service can be used and the argument once built can be imported to the workspace and associated with other elements of the task/activity instance. The recommendation service can be used to guide the worker along the whole process. All together, we can envisage a very efficient environment for production of informational elements in form and content compatible with the organization of the KWSP archive.

The set of services outlined above are far from exhaustive. We can conceive and design many more services for the KWSP platform. For example, we can think of building a service for *semantic comparison* of IEs. Another important service, *collaboration*, allowing multiple workers work together on a problem can also be easily built utilizing combined support of other services over a networked environment. We can also build specialized applications for performing one or more activities with respect to individual or a group of KWSSes to run on the top of the platform. For example we can conceive *transcription*, i.e., re-organizing the information contained in documents for putting them into KWSP archive, as a task and facilitate it through a KWSS. Now, a major activity in the task is for the worker to segment the document into informational element. However, we can use NLP-base document analysis techniques (Bestgen 2006) to partially automate the activity. Thus the tool used for document analysis can easily be identified with the transcription KWSS and the particular activity of the document analysis.

## TESTING THE DESIGN METHOD

According to Walls et al. (1992), testing the design method involves verifying "… whether the design method results in an artifact which hypotheses is consistent with the meta-design." In the following we shall examine the capabilities of the *design method* in realizing the artifacts of the meta-designs. Here, we again use the method of descriptive and reasoned arguments (Hevner et al. 2004).

In the KWSP platform, the KWSS definitions are used for describing activity structures and informational structures of the supported task types as well as the related domain vocabulary. They realize the KWSS-specific guidance/description artifacts (MDA 1) identified as a meta-design element. The design method accommodates for development and maintenance of these artifacts as well as their archival in the KWSP archive. Further, in KWSP, the systematic access and utilization of these artifacts is facilitated through various services, especially the workspace service.

The connectivity or association artifacts (MDA 2) are realized in KWSP by means of the design of generic informational element containers as objects capable of maintaining *links* with other elements. The KWSP, through the contextualized articulation service, facilitate creation of the links, which are archived along with the IEs. Other services exploit the links to provide various sophisticated system functionalities for supporting the knowledge workers.

In KWSP, creation and management of granular IEs (MDA 3), according to the definitions, with rich linkage reflecting support and provenance is provided through the contextualization support service. While archival of the same is achieved in the KWSP archive, the exploration service allows for retrieval of IEs from the archive based on sophisticated IR techniques.

As shown above, the KWSP platform and the KWSSes built on top of it together create an integrated and versatile environment for creation, maintenance and utilization of the design artifacts. The environment supports the knowledge workers to efficiently seek and consume relevant information using *exploration service* from the archive. The *contextualized articulation* or simply *articulation service* aids the workers to produce new information in a contextualized form. The *argumentation* service help workers to put together systematically complex information developed through creative and reflective efforts of the workers. Finally, the workspace service, configured according to the task-specific definitional elements, presents to the workers system interfaces customized for respective task types. Also the context-aware recommendation service helps the workers throughout the performance of their tasks. Thus the KWSP as a whole can be considered as the realization of the MDA 4.

It is easy to see that the environment can be further enriched with introduction of more services. Also, in this environment, creation and archival of new information is integral to process of performing knowledge works. Thus, with the continued use of the KWSSes, the platform gets enriched with accumulation more and more properly contextualized informational contents in the KWSP archive.

## DISCUSSIONS

In the process of developing the TOKM, we have discussed the major functionalities of KWSP and KWSS and the resulting benefits which can be enjoyed by the user community. However, the platform, as it is outlined above, can offer several other additional benefits of significant value. Here we briefly discuss a few of them.

Though we are not examining the behavioral aspects of the workers in TOKM, one of them becomes apparent. Since each individual KWSS support a particular task/task-type, its user community typically form a "Community of Practices" or CoP (Wenger 1998). A CoP is characterized by relative homogeneity in expertise, interests and sharing of stake in the quality performance of their tasks. This works in favor of the KWSS designers in studying the requirements of the task-type for developing the definitional elements.

As mentioned earlier, the platform offers opportunities of immensely valuable but difficult to achieve *double loop learning* (Argyris and Schön 1978). Here we outline some of the ways to achieve it in KWSP/KWSS. In KWSP both the definitional elements describing the best possible way, as perceived by the organization, of performing a task, i.e., the *espoused theory of work*, as well as the detail information on how actually the tasks instances are performed by the worker, i.e., the *theory at-work*, are available. Thus, careful analysis and comparison of them can unearth useful hints on how to improve the task performances by modifying the definitional elements to improve work practices.

In modern organizations many complex IT applications are used, many of which incorporate *expert knowledge* (e.g. rule base in an expert system) about the target problems and their solutions. However, as discussed by Vicente (2000) in context of financial computing, validity of such expert knowledge does not remain the same over time. Failure to maintain the accuracy of the systems with changing social environment can lead to disastrous results. Though no in-depth study is available, many trade publications point to such failures in many financial institutions as being among significant causes of current "global recession" we are going through. However, maintenance of knowledge bases are arduous tasks (Menzies 1999), some of it are not even computable (Debenham 2003).

Now, let us consider the problem from the TOKM perspective. Creating a knowledge-base is undoubtedly among most knowledge-intensive tasks. Thus, following TOKM, we can design a KWSS for the purpose, where the information regarding all the facts, assumptions, reasoning etc. used in building the knowledge-base can be articulated, recorded and archived systematically. The KWSS can also record how they are encoded in the knowledge base. Thus, any time in future, this information can be easily accessed and compared with environment of the time. This will allow us to estimate the validity of the knowledge base at that time and implement proper modifications, if required. Here also action theory (Argyris and Schön 1978) based approaches can be employed for performing the maintenance of such systems.

There are many other possible ways of exploiting the class of systems proposed here. For example, in large organizations, finding people with required expertise is one of the major problems (McDonald and Ackerman 1998). It is identified as an important part of one of the knowledge re-used types defined by Markus (2001). Various methods, including network analysis and text analysis, often in combination (Ehrlich et al. 2007) are used for this purpose. It can be easily seen that the way information is organized in the KWSP archive; they can be analyzed to find experts based on the objective information about their performance of various task instances. Our future research works will include investigating the research issues identified earlier as well as exploring innovative ways in which the powerful architecture of KWSP can be exploited.

# FUTURE RESEARCH DIRECTIONS

The TOKM opens up quite a few new but focused directions of further research. Some of them are already pointed out. While I plan to pursue a few of them, especially in the area of advanced exploration services, currently I am engaged in building a prototype KWSP and a patient-care KWSS. Currently various scenario-based testing of the prototype is being carried out. In near future, I plan to launch a pilot project with relatively small number of KWSSes (say, 3-4) in a real organizational environment. This will allow me study closely various aspects of the system such as scaling, suitable user-interface design and development of methodologies for building KWSS. It is apparent that realizing full potential of the TOKM is a large and long term enterprise. I hope that fellow researchers may merit the problems and possibilities pointed out in this paper worth their attentions and contribute towards full exploitation of their potentials. Possibly in future these systems will fit the bill of which are popularly known as "Next Generation Information Systems".